\documentclass[floats,prd,aps,amssymb,nofootinbib,twocolumn]{revtex4}
\usepackage{amssymb}
\usepackage{graphics}
\usepackage{amsmath}
\usepackage{amsfonts}
\usepackage{bm}
\usepackage[]{latexsym}
\usepackage{cellspace}
\usepackage{mathtools}
\usepackage{amstext}
\usepackage{amssymb}
\usepackage{stmaryrd}
\usepackage{stackrel}
\usepackage{graphicx}
\usepackage{esint}
\usepackage[utf8]{inputenc}
\usepackage{blindtext}
\usepackage{float}
\restylefloat{table}
\usepackage{booktabs}
\usepackage{enumitem} 
\usepackage{amssymb,amsbsy,epsfig,color,graphicx}

\usepackage{etoolbox} 
\usepackage{lipsum} 
\usepackage[capitalize]{cleveref}

\newcommand{\be}{\begin{equation}}\newcommand{\ee}{\end{equation}}
\newcommand{\bea}{\begin{eqnarray}}\newcommand{\eea}{\end{eqnarray}}
\newcommand{\brr}{\begin{array}}\newcommand{\err}{\end{array}}
\newcommand{\bit}{\begin{itemize}}\newcommand{\eit}{\end{itemize}}
\newcommand{\ben}{\begin{enumerate}}\newcommand{\een}{\end{enumerate}}

\newcommand{\ba}{\begin{array}}
\newcommand{\ea}{\end{array}}

\def\al{\alpha}

\def\1{{_{1}}}\def\2{{_{2}}}

\def\noHe0{:\;\!\!\;\!\!:H_e(0):\;\!\!\;\!\!:}
\def\noHm0{:\;\!\!\;\!\!:H_\mu(0):\;\!\!\;\!\!:}
\def\al{\alpha}

\def\1{{_{1}}}\def\2{{_{2}}}

\begin{document}
	
\title{Generalized Uncertainty Principle and Corpuscular Gravity}

\author{Luca Buoninfante\footnote{lbuoninfante@sa.infn.it}$^{\hspace{0.3mm}1,2,3}$, Giuseppe Gaetano Luciano\footnote{gluciano@sa.infn.it}$^{\hspace{0.3mm}2}$ and Luciano Petruzziello\footnote{lpetruzziello@na.infn.it}$^{\hspace{0.3mm}1,2}$}
 
\affiliation
{\vspace{1mm}$^1$Dipartimento di Fisica, Universit\'a di Salerno, Via Giovanni Paolo II, 132 I-84084 Fisciano (SA), Italy.
\\ 
\vspace{0mm}
$^2$INFN, Sezione di Napoli, Gruppo collegato di Salerno, Italy,
\\
\vspace{1mm}
$^3$Van Swinderen Institute, University of Groningen, 9747 AG, Groningen, The Netherlands.}

  \def\be{\begin{equation}}
\def\ee{\end{equation}}
\def\al{\alpha}
\def\bea{\begin{eqnarray}}
\def\eea{\end{eqnarray}}

\begin{abstract}
We show that the implications of 
the generalized uncertainty principle (GUP) in the black 
hole physics are consistent with the predictions of the
corpuscular theory of gravity, in which a black hole is 
conceived as a Bose--Einstein condensate 
of weakly interacting gravitons stuck 
at the critical point of a quantum phase transition. 
In particular, we prove that the GUP--induced shift of the
Hawking temperature can be reinterpreted in terms of 
non--thermal corrections to the spectrum of the black hole radiation, 
in accordance with the corpuscular gravity picture.
By comparing the two scenarios, we are able to estimate 
the GUP deformation parameter $\beta$, which turns out to be 
of the order of unity, in agreement with the expectations of some 
models of 
string theory. We also comment on the sign of 
$\beta$, exploring the possibility of having a negative deformation 
parameter when a corpuscular quantum description of the 
gravitational interaction is assumed to be valid.
\end{abstract}

 \vskip -1.0 truecm
\maketitle

\section{Introduction}

One of the most difficult and stimulating 
challenges the physics community has 
been struggling with for a long time is to 
understand whether the gravitational interaction 
has an intrinsic quantum nature and, if so, how 
to formulate a thorough quantum theory of gravity 
which avoids conceptual problems and is able 
to make successful predictions at any energy scale.
In pursuing the aim of combining both 
gravitational and quantum effects, the question inevitably arises  
as to whether the basic principles of quantum 
mechanics need to be revised in the quantum gravity
realm.

It is well known that one of the fundaments
of quantum mechanics is the Heisenberg Uncertainty 
Principle (HUP)\footnote{Throughout the paper, we work with the units $c=k_{\rm B}=1$, where $k_{\rm B}$ is the Boltzmann constant, and $\hbar\neq 1$. The Planck length is defined as $\ell_p=\sqrt{\hbar\,G}$, while the Planck mass as $m_p=\hbar/\ell_p$.}~\cite{aHeisenberg:1927zz},
\begin{equation}
\delta x\,\delta p\geq \frac{\hbar}{2}\,,\label{hup}
\end{equation}
which can be derived from the non--vanishing commutator 
between the position and momentum operators $\hat x$ and $\hat p$,
respectively, 
\begin{equation}
[\hat{x},\,\hat{p}]= i\hbar\,.
\label{comm-hup}
\end{equation}
The above inequality asserts that, in the quantum regime, 
the more precisely the position (momentum) of a particle is known, 
the less precisely one can say what its momentum (position) is.
This implies the existence of a region $\delta x\delta p$ 
of size $\hbar$ in the phase space, 
in which any physical prediction cannot 
be tested. In spite of this, however, no quantum limit
on sharp measurements of either position or momentum separately is
fixed a priori: in other terms, arbitrarily short distances may in principle be
detected via arbitrarily high energy probes, and vice--versa. 

The situation becomes more subtle if one tries to
merge quantum and gravity effects within the same framework. 
In that case, indeed, several models of quantum gravity
propose the existence of a minimum length at Planck scale~\cite{Maggiore:1993zu}
that accounts for a limited resolution of spacetime.
The Planck length $\ell_p$ thus appears as 
a natural threshold beyond which spacetime would no longer be smooth, 
but rather it would have a foamy structure due to inherent quantum fluctuations~\cite{Wheeler:1955zz}.

Along this line, many studies~\cite{Snyder:1946qz,Yang:1947ud,maed,Karolyhazy:1966zz,Amati:1987wq,Gross:1987kza,Amati:1988tn,Konishi:1989wk,Maggiore:1993kv,Scardigli:1995qd,FS,Capozziello:1999wx,Bojowald:2011jd,Scardigli:2003kr,Adler:1999bu,Adler,Casadio:2009jc,Casadio:2013aua,Scardigli:2016pjs,VAF,Lambiase:2017adh,Scardigli:2018jlm,Luciano2019} 
have converged on the idea that the HUP should be 
properly modified at the quantum gravity scale, in order 
to accommodate the existence of such a fundamental length. 
In this sense, one of the most adopted  generalizations of the 
uncertainty principle (GUP) reads
\begin{equation}
\delta x\,\delta p\ge \frac{\hbar}{2}\pm2\hspace{0.3mm}|\beta|\hspace{0.3mm}\ell_p^2\frac{\delta p^2}{\hbar}=\frac{\hbar}{2}\pm2\hspace{0.3mm}|\beta|\hspace{0.3mm}\hbar\frac{\delta p^2}{m_p^2}\,,
\label{gup}
\end{equation}
where the sign $\pm$ refers to positive/negative
values of the dimensionless \emph{deformation parameter} $\beta$, 
which is assumed to be of order unity
in some models of quantum gravity, and in particular in string theory~\cite{Amati:1987wq,Amati:1988tn}. However, one could also investigate Eq.~\eqref{gup} 
from a phenomenological point of view, and seek experimental bounds
on $\beta$ (for a recent review of the various phenomenological approaches, 
see, for example, Ref.~\cite{Kanazawa:2019llj}). 
Clearly, for $\beta\hbar/m_p^2\rightarrow0$, standard quantum mechanics is
recovered,  so that modifications to the HUP
only become relevant at the Planck
scale, as expected. 
Moreover, for mirror--symmetric states 
(i.e. $\left\langle \hat{p}\right\rangle =0$), Eq.~\eqref{gup} can be
deduced from the modified commutator 
\begin{equation}
[\hat{x},\,\hat{p}]= i\hbar\left(1\pm|\beta|\left(\frac{\hat{p}^2}{m_p^2}\right)\right).\label{comm-gup}
\end{equation}

If, on the one hand, the assumption that $\beta\sim\mathcal{O}(1)$
is quite generally accepted and it has also been confirmed 
by achievements in contexts other than string theory~\cite{Scardigli:2016pjs,VAF,Lambiase:2017adh,Capozziello:1999wx,Luciano2019}, 
on the other hand the problem of the sign of $\beta$ is much more debated.
Although in various derivations and gedanken 
experiments on GUP it seems more reasonable
to have a positive parameter, arguments in favor of the opposite choice
are not lacking. For instance, in Ref.~\cite{Jizba:2009qf} it was emphasized 
that the GUP with $\beta<0$ would be consistent with a description in which
the universe has an underlying crystal lattice--like structure.
Similarly, in Ref.~\cite{Ong:2018zqn} a 
negative GUP parameter was proved to be 
the only setting compatible with the 
Chandrasekhar limit for white dwarfs. A further confirmation
was given in Ref.~\cite{Kanazawa:2019llj} in the context
of non--commutative Schwarzschild geometry.

In this connection, a clue to an answer for such
a dichotomy may be found
within the framework of Corpuscular Gravity (CG) originally formulated 
in Refs.~\cite{Dvali:2011aa,Dvali:2012en}, 
and then revisited from a complementary point of view in Refs.~\cite{Casadio:2013ulk,Casadio:2014vja} (see also Refs.~\cite{Dvali:2013vxa,Dvali:2013eja,Dvali:2014ila,Dvali:2015aja,Casadio:2015bna,Casadio:2016aum,Casadio:2016zpl,giusti-tesi} for further developments).
According to this model, black holes can be understood as 
a Bose--Einstein condensate of weakly interacting 
gravitons at the critical
point of a quantum phase transition. 
As a consequence of this portrait, black hole characteristics
get non--trivially modified; for instance, one can show
the evaporation rate 
gains a correction of order $1/N^{3/2}$ with respect to
the standard semiclassical expression, where $N\sim{(M/m_p)}^2$ is
the number of constituents (gravitons). 

To the best of our knowledge, the GUP and CG approaches to
black hole physics 
have been regarded as completely unrelated treatments
to date; 
indeed, whilst corrections arising 
in the former case are typically assumed to be thermal, 
and, therefore, can be cast in the 
form of a shift of the Hawking temperature~\cite{Adler}, the effects induced
by the CG description do spoil the thermality of the
black hole radiation~\cite{Dvali:2011aa}, giving
rise to seemingly different pictures.

Starting from the outlined scenario, in the present paper
we provide a link between the GUP black hole thermodynamics 
and the corpuscular model of gravity, showing 
that GUP corrections to the 
Hawking temperature can be equivalently rephrased in terms of
a non--thermal distortion of the spectrum.
This allows for a straightforward comparison of 
GUP and CG predictions. In particular, we infer that 
the two approaches are consistent, provided
that $\beta$ is of order unity, 
in agreement with string theory.
We further comment on the sign of $\beta$, speculating on
the possibility to obtain a negative value.

The paper is organized as follows: Section~\ref{uncert-sec}
is devoted to a review of black hole thermodynamics in the context of 
the GUP. 
We compute corrections to the Hawking temperature
and the evaporation rate, both for positive and negative 
values of the deformation parameter $\beta$. Then, we show
how to recast the ensuing shift of the temperature in terms on 
a non--thermal spectrum.
In Section~\ref{corpusc-sec} we discuss some fundamental 
aspects of CG, focusing on the
derivation of the formula for the emission rate. 
GUP and CG approaches are then compared in
Section~\ref{comparis-sec}. By requiring that
the modified expressions of the evaporation rate derived in the
two frameworks are equal  (at the
first order), we are able to evaluate the GUP parameter
$\beta$. Finally, conclusions and discussion can be found in Section~\ref{conclus-sec}.


\section{Uncertainty relations and black hole thermodynamics}\label{uncert-sec}

Let us consider a spherically symmetric black hole with mass $M$ and 
Schwarzschild radius $r_s=2MG$. Following the 
arguments of Refs.~\cite{Scardigli:1995qd,Adler}, 
the Hawking temperature $T_{\rm H}$  
can be derived in a heuristic
way by using the HUP~\eqref{hup} and general
properties of black holes.
To this aim, let us observe that,  
just outside the event horizon, 
the position uncertainty of photons
emitted by the black hole
is of the order of its Schwarzschild radius, i.e. $\delta x\simeq \mu\hspace{0.2mm}r_s$, where 
the constant $\mu$ is of order of unity
and will be fixed below. From Eq.~\eqref{hup}, the corresponding
momentum uncertainty is given by
\be
\label{momunc}
\delta p\simeq \frac{\hbar}{4\mu\hspace{0.2mm}M\hspace{0.2mm}G}, 
\ee
which also represents the characteristic
energy of the emitted photons, since $\delta p\simeq p=E$.
According to the equipartition theorem,
this can be now identified with the temperature $T$
of the ensemble of photons,
\be
\label{temp}
E=T\simeq \frac{\hbar}{4\mu\hspace{0.2mm}M\hspace{0.2mm}G}=\frac{m^2_p}{4\mu\hspace{0.2mm}M}
\ee
which agrees with the Hawking temperature, 
\be
\label{HT}
T_{\rm H}=\frac{\hbar}{8\pi\hspace{0.2mm}M\hspace{0.2mm}G}\equiv\frac{m_p^2}{8\pi M} 
\ee 
provided that $\mu=2\pi$. 

Therefore, on the basis of the HUP and
thermodynamic consistency, 
we have recovered the standard Hawking formula Eq.~\eqref{HT}
for the temperature of the radiation emitted by the black hole.

Now, it is well known that  black holes with temperature greater 
than the background temperature (about $2.7\,\mathrm{K}$ for the
present universe) shrink over time
by radiating energy in the form of photons and
other ordinary particles. In certain conditions~\cite{Page:1976df}, however, 
it is reasonable to assume that the evaporation  
is dominated by photon emission. In this case, we can exploit
the Stefan--Boltzmann law to estimate the radiated power $P$ as
\begin{equation}
\label{rate}
P=A_s\hspace{0.3mm}\varepsilon\hspace{0.3mm}\sigma\,T^4\simeq A_s\hspace{0.3mm}\sigma\,T^4,
\\[2mm]
\end{equation}
where $A_s=4\pi\hspace{0.2mm}r_s^2$ is the black hole sphere surface area 
at Schwarzschild radius $r_s$, $\sigma=\pi^2/60\hbar^3$ is the 
Stefan--Boltzmann constant, and we have assumed for simplicity
the black hole to be a perfect blackbody, i.e. $\varepsilon\simeq 1$.

Using Eqs.~\eqref{temp} and~\eqref{rate}, the black hole
energy loss can be easily evaluated as a function of time, yielding
\be
\label{bhel}
\frac{dM}{dt}=-P\simeq-\frac{1}{60\hspace{0.3mm}{(16)}^2\hspace{0.2mm}\pi\hspace{0.3mm}\sqrt{\hbar\hspace{0.3mm}G}}\,\frac{m_p^3}{M^2}=-\frac{1}{60\hspace{0.3mm}{(16)}^2\hspace{0.2mm}\pi}\,\frac{m_p^4}{\hbar\, M^2}.
\ee
Therefore, the evaporation process
leads black holes to vanish entirely with
both the temperature~\eqref{temp} and emission rate~\eqref{bhel} blowing up as the mass decreases.

The above results have been derived starting from the 
HUP in Eq.~\eqref{hup}. We now wish to follow a similar procedure by resorting to
Eq.~\eqref{gup}, so as to realize to what
extent the GUP affects the black hole thermodynamics.
In this case, solving Eq.~\eqref{gup} with respect
to the momentum uncertainty $\delta p$ and setting again $\delta x$ 
of the order of the Schwarzschild radius, we obtain the following
expression for the modified Hawking temperature
\begin{equation}
\label{guptuniversal}
T_{\rm GUP}=\pm\frac{M}{4\pi|\beta|}\left(1\,\pm\,\sqrt{1\,\mp\,|\beta|\frac{m^2_p}{M^2}}\right).\\[2mm]
\end{equation}
In the semiclassical limit $\sqrt{|\beta|} m_p/M\ll 1$, this agrees
with the standard Hawking result in Eq.~\eqref{temp}, provided that
the negative sign in front of the square root is chosen, whereas 
the positive sign has no physical meaning.
Similarly, the emission rate in Eq.~\eqref{bhel}
is modified as
\be
\label{newrate}
\hspace{-1.5mm}
\left(\frac{dM}{dt}\right)_{\rm \hspace{-1mm}GUP}\simeq\frac{-1}{60\, (16)\,\pi\,{\hbar}}\,\frac{M^6}{{|\beta|}^4\,m_p^4}{\left(1-\sqrt{1\mp|\beta|\,\frac{m_p^2}{M^2}}\right)}^{\hspace{-0.5mm}4}.
\ee
In what follows, the implications of Eqs.~\eqref{guptuniversal}
and \eqref{newrate}
will be discussed separately for the cases of 
$\beta>0$ and $\beta<0$.

\subsection{GUP with $\beta>0$}
Let us start by analyzing the most common
setting of GUP with positive deformation parameter. 
In this case, from Eq.~\eqref{guptuniversal} it is easy to see that the GUP 
naturally introduces a minimum size 
allowed for black holes: for $M<\sqrt{\beta}\,m_p$ (i.e. 
$r_s<2\sqrt{\beta}\,\ell_p$), indeed, the temperature would
become complex. This means that the evaporation
process should stop at $M\sim \sqrt{\beta}\,m_p$ ($r_s\sim\sqrt{\beta}\,\ell_p$), thus
leading to an inert remnant with finite temperature and size~\cite{Adler}, in contrast
with predictions of ordinary black hole thermodynamics.
We remark that the idea of black hole remnants
dates back to Aharonov--Casher--Nussinov, 
who first addressed the issue in the context of the 
black hole unitarity puzzle~\cite{Aharonov:1987tp}.

Similarly, concerning the modified emission rate Eq.~\eqref{newrate},
we find that it is finite at the endpoint of black hole evaporation $M\sim\sqrt{\beta}\,m_p$, 
whereas the corresponding HUP result~\eqref{bhel} diverges   
at the endpoint when $M=0$. Again, we stress that the standard 
behavior in Eq.~\eqref{bhel} is recovered for $\sqrt{\beta} m_p/M\ll1$.

\subsection{GUP with $\beta<0$}
Although the GUP with $\beta>0$ cures the 
undesired infinite final temperature predicted
by Hawking's formula~\eqref{HT} giving rise
to black hole remnants, it would create several 
complications, such as the entropy/information
problem~\cite{Mathur:2009hf,Bekenstein:1993dz}, or the removal of
the Chandrasekhar limit~\cite{Ong:2018zqn}. 
The latter prediction, in particular, would allow white dwarfs to 
be arbitrarily large, a result that is at odds with astrophysical observations.
An elegant way to overcome these ambiguities 
was proposed in Ref.~\cite{Ong:2018zqn}, 
where it was shown that both the infinities in 
black hole and white dwarf physics can be
avoided by choosing a negative deformation parameter in Eq.~\eqref{gup}.
A similar scenario had previously been
encountered in Ref.~\cite{Jizba:2009qf}
in the context of GUP in a crystal-like universe
with lattice spacing of the order of Planck length.

Let us then consider the case $\beta<0$. 
With this setting, from Eqs.~\eqref{bhel} and~\eqref{newrate} 
we  obtain that both the modified temperature and emission rate
are  well defined even for $M<\sqrt{|\beta|}\,m_p$. 
For a sufficiently small $M$, in particular, 
the modified temperature in Eq.~\eqref{guptuniversal} can be approximated as
\be
\label{modTexp}
T_{\rm GUP}\simeq \frac{m_p}{4\hspace{0.2mm}\pi\hspace{0.2mm}\sqrt{|\beta|}}<\infty.
\ee
Although no lower bound on the black hole size arises in this
framework, the Hawking temperature remains finite as the black hole
evaporates to zero mass. 
From Eq.~\eqref{modTexp} we also deduce that
the bound on the Hawking temperature
is independent of the initial black hole mass.

\subsection{GUP induced non--thermal corrections}
So far, we have assumed that the correction induced 
by the GUP has a thermal character, and, therefore,
it can be cast in the form of a shift of the Hawking temperature which
does not affect the overall thermality of the Planck spectrum 
of the black hole, i.e.
\begin{equation}
\mathcal{N_{\rm{H}}}(\omega)=\frac{1}{e^{\hbar \omega/T_{\rm H}}-1}\,\underset{\scriptscriptstyle{{T\rightarrow T_{{}_{\scriptscriptstyle{\rm GUP}}}}}}{\longrightarrow}\,
\mathcal{N}_{\rm GUP}(\omega)=\frac{1}{e^{\hbar \omega/T_{\rm GUP}}-1}.\label{GUP-distrib}
\end{equation}
Here, we show that such
a viewpoint can be safely reversed.
In other terms, starting from the above results, 
we reinterpret GUP effects on the black hole
radiation as corrections that spoil
the thermal nature of the  
Planck distribution, while leaving 
the Hawking formula~\eqref{HT} unchanged.
 
To this aim, let us observe that, by inserting
the modified Hawking temperature~\eqref{guptuniversal} in 
Eq.~\eqref{GUP-distrib} and expanding
for $\sqrt{|\beta|} m_p/M\ll 1,$ it follows that
\begin{equation}
\mathcal{N}_{\rm GUP}(\omega)\simeq\frac{1}{e^{\hbar \omega/T_{\rm{H}}}-1}\left(1\mp |\beta|f(\omega)\frac{m_p^2}{M^2}\right),\label{GUP-non-thermal}
\end{equation}
where $\mp$ corresponds to positive or negative deformation parameters, respectively, and we have defined
\begin{equation}
f(\omega)\equiv\frac{\hbar \omega}{4T_{\rm{H}}}\frac{e^{\hbar \omega/T_{\rm{H}}}}{e^{\hbar \omega/T_{\rm{H}}}-1}.
\end{equation}
Equation~\eqref{GUP-non-thermal} does provide the
expected result, 
since it states that the thermal 
distribution~\eqref{GUP-distrib} with a 
shifted temperature $T=T_{GUP}$
can be rewritten as the standard Planck factor with
$T=T_{\rm H}$, plus non-thermal corrections depending
on the GUP parameter $\beta$.

In what follows, we shall essentially refer to this
alternative interpretation 
of the GUP-induced correction. Notice that, although it is
quite unusual in the context of black hole
physics, a similar result has been recently derived
for the Unruh radiation within the framework of QFT
with modified commutation relations (see
Ref.~\cite{Scardigli:2018jlm} for details). In light of this,  
it is thus reasonable to 
expect that a non-thermal distortion of the spectrum
may be also derived
via a more rigorous QFT treatment of the 
black hole thermodynamics 
in the presence of the GUP.
We further remark that, by considering the 
non--thermal distribution~\eqref{GUP-non-thermal} 
for the black hole radiation, the Stephan--Boltzmann 
law~\eqref{rate} should  be accordingly modified, 
and this must be consistent with the relation~\eqref{newrate} 
for the evaporation rate.

In the next Section, it will be shown 
how the above interpretation is conceptually fundamental
to establish a correspondence
between the GUP and CG approaches to
black hole physics.

\section{Corpuscular black holes}\label{corpusc-sec}

It is well known that, in general relativity, black holes are 
solutions of the Einstein field equations that are fully characterized by means of three parameters only (no--hair theorem)~\cite{hawk}: mass, charge and angular momentum. This suggests that, classically, a black hole can carry a little amount of information. 
Conversely, from the quantum mechanical point of view, a black hole possesses a huge number of states due to its extremely large entropy, and this is the cause that allows for the emergence of the information paradox~\cite{Mathur:2009hf}. Indeed, if we perform the classical limit starting from such a quantum picture, it appears that the entropy is infinite (i.e. the number of states is infinite), but this is in contradiction with the fact that a classical black hole is featureless, according
to what discussed above~\cite{Dvali:2018tno}.

The aforementioned paradox can be elegantly solved within the framework of Corpuscular Gravity (CG), which was introduced in Refs.~\cite{Dvali:2011aa,Dvali:2012en}, and revised from a different perspective in Refs.~\cite{Casadio:2013ulk,Casadio:2014vja,Dvali:2013vxa,Dvali:2013eja,Dvali:2014ila,Dvali:2015aja,Casadio:2015bna,Casadio:2016aum,Casadio:2016zpl,giusti-tesi}.
According to this picture, indeed, black holes 
can be conceived as Bose--Einstein condensates of $N$ interacting and non--propagating longitudinal gravitons, and thus as intrinsically quantum objects\footnote{Let us recall that a Bose--Einstein condensate of attractive bosons and their phase transitions were firstly studied in Ref.~\cite{Bogolyubov:1947zz} in the context of condensed matter physics.}. As argued in Ref. \cite{Dvali:2015aja}, the corpuscular nature of the gravitational interaction induces non--thermal corrections to the black hole radiation, which scale as $1/N.$ The thermal Hawking radiation of the semiclassical picture is then consistently recovered in the limit $N\rightarrow \infty.$  

In order to better understand such a black hole quantum's portrait and
make a comparison with the GUP predictions, 
let us review the most relevant properties of 
the corpuscular theory of gravity.

\subsection{Bose-Einstein condensate of gravitons}

Let us consider a Bose--Einstein condensate of total mass $M$ and radius $R$, which is made up of $N$ weakly interacting gravitons. At low energy, we can define a quantum gravitational self--coupling for each single graviton of wavelength $\lambda$ as follows~\cite{Dvali:2011aa}
\begin{equation}
\alpha_g\equiv\frac{\hbar G}{\lambda^2}=\frac{{\ell}_p^2}{\lambda^2}\,. \label{self-coup}
\end{equation}
One of the main features of a Bose--Einstein condensate is that, due to the interaction, its constituents acquire a collective behavior, so that
their wavelengths get increasingly  larger and their masses smaller;
strictly speaking, the constituents become {\it softer} bosons. In particular, most of the gravitons composing the gravitational system will have a wavelength of the order $\lambda\sim R$, namely of the order of the size of the system itself\footnote{More precisely, the condensate is also made up of gravitons with shorter wavelengths, $\lambda<R.$ However, all physical quantities will be mostly determined by those gravitons that possess a wavelength of the order $\lambda\sim R$, since the contribution arising from \emph{hard} gravitons turns out to be exponentially suppressed and can thus be neglected~\cite{Dvali:2011aa}.}. Hence, similarly to Eq.~\eqref{self-coup}, it is possible to define a collective quantum coupling as 
\begin{equation}
N\alpha_g\equiv N\frac{\hbar G}{\lambda^2}\simeq N\frac{{\ell}_p^2}{R^2}. \label{coll-coup}
\end{equation}
We now seek the relation that links the total mass of a Bose--Einstein condensate and its radius to the number $N$ of quanta composing the system. By performing a standard computation, one can show that the gravitational binding energy of the system is given by
\begin{equation}
E_g\simeq \frac{GM^2}{R}. \label{grav-energ}
\end{equation}
On the other hand, from a purely quantum point of view, the binding energy can be expressed as the sum of the energies associated to each single graviton, i.e.\footnote{Note that, in Eq.~\eqref{quantum-grav-energ}, we are allowed to employ the usual de Broglie relation because we are considering low energy weakly interacting gravitons with large wavelengths. Quantum gravity effects only emerge as a collective feature of the whole gravitational system.}
\begin{equation}
E_g\simeq N\frac{\hbar}{\lambda}\simeq N\frac{\hbar}{R}. \label{quantum-grav-energ}
\end{equation}
Therefore, by comparing Eqs.~\eqref{grav-energ} and \eqref{quantum-grav-energ}, we obtain 
\begin{equation}
M\simeq \sqrt{N}m_p, \label{total mass}
\end{equation}
which also implies for the Schwarzschild radius 
\begin{equation}
r_{s}\simeq \sqrt{N}{\ell}_p. \label{sch-radius-N}
\end{equation}
By assuming that the size of the condensate is $R\sim r_s$ (i.e. the overall gravitational system is a black hole) and using the expression in Eq.~\eqref{sch-radius-N} for the Schwarzschild radius, we notice that the collective quantum coupling defined in Eq.~\eqref{coll-coup} is always of order unity in the case of a black hole
\begin{equation}
N\alpha_g\simeq 1. \label{coll-coup=1}
\end{equation}
In condensed matter physics, it is well known that the inequality $N\alpha_g<1$ corresponds to a phase in which a Bose--Einstein condensate is weakly interacting. On the other hand, the equality $N\alpha_g=1$ represents a critical point at which a phase transition occurs, thus letting the condensate become strongly interacting, whereas for $N\alpha_g>1$ it is possible to observe only a strongly interacting phase~\cite{Bogolyubov:1947zz}. 
Thus, in this quantum corpuscular picture, a black hole can be defined as a Bose--Einstein condensate of gravitons 
stuck at the critical point of a quantum phase transition~\cite{Dvali:2012en}. This is a crucial property which ensures that a gravitational bound system can exist for any $N$ or, in other words, that such a graviton condensate is self--sustained.

\subsection{Thermodynamic properties of corpuscular black holes}

We now analyze some thermodynamic aspects of quantum corpuscular black holes, and in particular we show that gravitons can escape from the
considered system. Such a phenomenon represents the corpuscular counterpart of the black hole radiation emission~\cite{Dvali:2011aa}. 

First of all, we need to compute the probability for a graviton to escape from a gravitational bound state, namely we have to determine the so--called \emph{escape energy} and \emph{escape wavelength} of a single graviton. To this aim, observe that, for $N$ weakly interacting quanta composing a condensate of radius $R$ and mass $M$, a quantum gravitational interaction strength can be defined as~\cite{Dvali:2011aa}
\begin{equation}
\hbar N\alpha_g\equiv\hbar N\frac{L_p^2}{\lambda^2}, \label{inter-strength}
\end{equation}
so that each graviton is subject to the following binding potential
\begin{equation}
E_{\rm esc}=\frac{\hbar N\alpha_g}{R}, \label{coll-potential}
\end{equation}
which is the threshold to exceed in order to escape. The corresponding escape wavelength is defined as
\begin{equation}
E_{\rm esc}=\frac{\hbar}{\lambda_{\rm esc}}\,. \label{de-broglie-escape}
\end{equation}
If we now employ Eqs.~(\ref{sch-radius-N}) and (\ref{coll-coup=1}) for the case of a black hole, we obtain
\begin{equation}
E_{\rm esc}\simeq \frac{\hbar}{\sqrt{N}\ell_p}\,,\quad\,\,\lambda_{\rm esc}\simeq \sqrt{N}\ell_p\,. \label{escape}
\end{equation}
This means that, although $N$ gravitons of wavelength $\lambda\sim~\hspace{-2mm}\sqrt{N}\ell_p$ can form a gravitational bound state, at the same time a depletion process is present, which is traduced in a leakage of the constituents of the condensate for any $N$. Clearly, this is related to the fact that $\lambda_{\mathrm{esc}}$ coincides with the wavelength of each graviton belonging to the condensate, that is, $\sqrt{N}\ell_p.$ In this sense, a black hole is a leaky Bose--Einstein condensate stuck at its critical point.

In terms of scattering amplitudes, the above picture can be regarded as a $2\rightarrow 2$ scattering process, in which one of the two gravitons is energetic enough to be able to exceed the threshold given by $E_{\rm esc}$. 

We can also obtain an estimation for the depletion rate $\Gamma$ of such a process. As usual, this should be given by a product involving the squared coupling constant $\alpha_g^2$, the characteristic energy scale of the process $E_{\rm esc}$ and a combinatoric factor $N(N-1)$, which can be approximated by $N^2$ for a very large number of constituents~\cite{Dvali:2011aa}, i.e.
\begin{equation}
\Gamma\simeq \alpha_g^2 N^2E_{\rm esc}\simeq \frac{\hbar}{\sqrt{N}\ell_p}. \label{rate-gamma}
\end{equation}
From the above relation, we can easily obtain the corresponding time scale of the considered process, which is given by $\Delta t=\hbar/\Gamma\simeq \sqrt{N}\ell_p.$

On the other hand, Eq.~\eqref{rate-gamma} allows us to infer
the mass decrease over time of the condensate, i.e.
\begin{equation}
\frac{dM}{dt}=-\frac{\Gamma}{\lambda_{\rm esc}}\simeq -\frac{\hbar}{N\ell_p^2}\simeq-\frac{m_p^4}{\hbar M^2}\,, \label{evaporation-rate-CG}
\end{equation}
which can be cast in terms of
the rate of emitted gravitons by use of Eq.~\eqref{total mass}, 
\begin{equation}
\frac{dN}{dt}\simeq -\frac{1}{\sqrt{N}\ell_p}. \label{evaporation-rate-N}
\end{equation}
We stress that, up to the factor  $1/[60(16)^2\pi]$, Eq.~\eqref{evaporation-rate-CG} reproduces the thermal evaporation rate of a black hole in Eq.~\eqref{bhel}, assuming the Hawking temperature in the corpuscular 
model to be given by~\cite{Dvali:2011aa}
\begin{equation}
T_{\rm {H}}\simeq \frac{\hbar}{\sqrt{N}\ell_p}\simeq\frac{m_p^2}{M}. \label{T_H-Corpuscul}
\end{equation}
We can also estimate the lifetime $\tau$ of a quantum black hole; indeed, by imposing $dN/dt\sim -N/\tau,$ we get
\begin{equation}
\tau\simeq  N^{3/2}\ell_p\simeq \frac{\hbar M^3}{m_p^4}\,.\label{life-time}
\end{equation}
For the sake of completeness, we also mention that the gravitational entropy $S$ of a corpuscular black hole takes the simple form $S\sim N.$ As a consequence, the number of states a black hole can use to store information is exponentially large, and it is given by $e^{N}$ \cite{Dvali:2011aa}.

\subsection{Non-thermal features of corpuscular black holes}

We have seen that the black hole quantum $N$-portrait manages to reproduce the semiclassical result, according to which a black hole emits a thermal radiation with temperature given by the Hawking formula~\eqref{HT}. However, from a more scrupulous investigation, one can see that such a result holds true only to the leading order, since in general there will be non--thermal corrections to the spectrum which scale as negative powers of the number of gravitons (see Refs.~\cite{Dvali:2011aa,Dvali:2012en,Dvali:2015aja} for a more detailed discussion on the non-thermal nature of the black hole radiation in CG and the ensuing resolution of the information loss paradox).

In this connection, notice that, in the computation of the depletion rate $\Gamma$ in Eq.~\eqref{rate-gamma}, we have only considered the simplest kind of interaction (i.e. a tree-level scattering diagram with two vertices); nevertheless, 
one expects that even higher-order processes provide $\Gamma$ with contributions that induce gravitons to escape. For instance, the next relevant $2\rightarrow2$ scattering process would possess three vertices, thus contributing with terms proportional to
$\alpha_g^3.$ Therefore, up to first order corrections, the depletion rate would take the form
\begin{equation}
\begin{array}{rl}
\Gamma\simeq& \displaystyle \alpha_g^2 N^2E_{\rm esc}+\mathcal{O}\left(\alpha_g^3 N^2E_{\rm esc}\right)\\[2.5mm]
\simeq & \displaystyle \frac{\hbar}{\sqrt{N}\ell_p}+\mathcal{O}\left(\frac{\hbar}{\ell_p}\frac{1}{N^{3/2}}\right). \label{rate-gamma-first-order}
\end{array}
\end{equation}
As for the leading order in Eq.~\eqref{evaporation-rate-CG}, 
the mass decrease of the black--hole can be now estimated from the modified
depletion rate~\eqref{rate-gamma-first-order}, obtaining~\cite{Dvali:2012en}
\begin{equation}
\frac{dM}{dt}\simeq -\frac{m_p^4}{\hbar M^2}+\mathcal{O}\left(\frac{m_p^6}{\hbar M^4}\right). \label{evap-rate-first-order}
\end{equation}
These last findings strengthen the awareness that, in the CG theory, black holes can be uniquely described through the variable $N$. Indeed, once the number of gravitons composing the condensate is known, we can determine all the macroscopic physical quantities (i.e. the mass, the radius, the evaporation rate, etc.). Clearly, this feature appears to hold also for the corrections to the semiclassical formulas, as Eqs.~(\ref{rate-gamma-first-order}) and (\ref{evap-rate-first-order}) denote.


\section{Consistency between GUP and corpuscular gravity}\label{comparis-sec}

In the previous Sections, the evaporation rate of a black hole 
has been computed within both the GUP and CG frameworks. 
Here, we compare the two expressions:
as it will be shown, this allows us to set
the value of the GUP deformation parameter $\beta$ for
which the GUP and CG treatments are consistent. 

For this purpose, we consider the GUP--modified expressions of the
emission rate in
Eq.~\eqref{newrate} 
expanded up to the order $\mathcal{O}(1/M^4)$, 
and the CG outcome in Eq.~\eqref{evap-rate-first-order} with the 
proper coefficients restored. This yields
\begin{equation}
\left(\frac{dM}{dt}\right)_{\rm GUP}\simeq \frac{-1}{60(16)^2\pi}\left(\frac{m^4_p}{\hbar M^2}\pm|\beta|\frac{m_p^6}{\hbar M^4}\right),
\label{ev-gup}
\end{equation}
and
\begin{equation}
\left(\frac{dM}{dt}\right)_{\rm CG}\simeq \frac{-1}{60(16)^2\pi}\left(\frac{m^4_p}{\hbar M^2}+\mathcal{O}\left(\frac{m_p^6}{\hbar M^4}\right)\right),
\label{ev-cg}
\end{equation}
where we recall that the sign $\pm$ in Eq.~\eqref{ev-gup} 
corresponds to a positive/negative value of the 
deformation parameter $\beta$. Note that, 
at least up to the leading order, the GUP- and 
CG-induced corrections turn out to have 
the same functional dependence on the black hole mass. Furthermore, since the
coefficient in front of the correction in Eq.~\eqref{ev-cg} is predicted
to be of order unity~\cite{Dvali:2011aa,Dvali:2012en}, numerical consistency between the two expressions automatically leads to 
%
\begin{equation}
|\beta|\sim \mathcal{O}(1),
\label{beta-1}
\end{equation}
which is in agreement with the predictions of other models of quantum gravity, and in particular with string theory~\cite{Amati:1987wq,Amati:1988tn}. 
Therefore, in spite of their completely different underlying backgrounds, the GUP and CG approaches are found to be compatible with each other.

However, the result~\eqref{beta-1} does not give any specific information about the sign of $\beta$. Since a full-fledged analytic derivation of Eq.~\eqref{ev-cg} including also higher-order scattering processes is still lacking, a definitive conclusion on this issue cannot be reached. On the one hand, 
relying on basic considerations on the nature of the scattering amplitudes, we would naively expect the second term in Eq.~\eqref{ev-cg} to contribute with the same sign as the leading order term, because we are only adding higher order diagrams describing the probability for a graviton to escape from the condensate. This would yield a positive value for the deformation parameter.

On the other hand, there are different claims which assert that the first order correction in the depletion rate~\eqref{rate-gamma-first-order} should be opposite to the leading order term, in such a way to slightly decrease the evaporation rate of the black hole given in Eq.~\eqref{ev-cg}. 
This was shown, for example,  within the framework of Horizon Quantum Mechanics~\cite{horizon-QM}, where
the depletion rate up to first order correction
takes the form~\cite{Casadio:2015bna}
\begin{equation}
\Gamma\simeq \frac{\hbar}{\sqrt{N}\ell_p}-\frac{3\hbar\gamma^2 N_H^2}{\ell_pN^{3/2}}\left(6\,\zeta(3)-\frac{\pi^4}{15}\right),
\label{deplet-horiz-qm}
\end{equation}
which would imply the following formula for the evaporation rate
\begin{equation}
\begin{array}{rl}
\displaystyle \left(\frac{dM}{dt}\right)_{\rm CG} \simeq& \displaystyle \frac{-1}{60(16)^2\pi}\left[\frac{m_p^4}{\hbar M^2}\right.\\[4mm]
&\hspace{-6mm}\displaystyle \,\,\,\,\,\,\,\,\,\,\left.-3\,\gamma^2 N_H^2\left(6\,\zeta(3)-\frac{\pi^4}{15}\right)\frac{m_p^6}{\hbar M^4}\right],
\label{evap-horiz-qm}
\end{array}
\end{equation}
where $N_H\equiv \sqrt{3}/\sqrt{\pi^2-6\zeta(3)}\simeq 1.06$ and $\zeta(x)$ is the Riemann zeta function. Note also that the constant factor in Eq.~\eqref{evap-horiz-qm} is given by $3\hbar\gamma^2N_H^2\left(6\zeta(3)-\pi^4/15\right)\simeq 2.7\gamma^2$ and $\gamma$ may be of order one~\cite{Casadio:2015bna}. 
With such a setting, the comparison of Eqs.~\eqref{ev-gup} and~\eqref{evap-horiz-qm} would further confirm the result  in Eq.~\eqref{beta-1}, but it would lead to a negative value for the deformation parameter, 
\be
\label{negative}
\beta<0\,.
\ee 
Moreover, we remark that  positive corrections to the evaporation rate of a black hole are required by the principle of energy conservation~\cite{Casadio:1997yv,Alberghi:2006km}, thus enforcing the validity of Eq.~\eqref{negative}. A more detailed discussion on the 
physical meaning and implications of  such a result can be found in the Conclusions.



\section{Conclusions}\label{conclus-sec}

In this paper, we have analyzed
to what extent the black hole thermodynamics gets
modified both in the presence of a generalized 
uncertainty principle (GUP) and in the corpuscular gravity (CG) theory. 
After remarking that, in both contexts,
corrections to the standard semiclassical results
can be viewed as originating from non-thermal 
deviations of the Hawking radiation, 
we have focused on the computation 
of the evaporation rate of a black hole.
By comparing the expressions derived within the
two frameworks,  we have finally managed to 
estimate the GUP deformation parameter $\beta$.
Specifically, in order for the
GUP and CG predictions to be consistent, 
we have found that
$\beta$ must be of order unity. 
This is certainly a non-trivial result, since
it states that the GUP and CG approaches
to black hole physics are
consistent with each other, and also 
with some other models of quantum gravity
such as string theory.

Furthermore, we have speculated on the sign of $\beta$.
Although  on this matter we are still far from
the definitive solution, 
a preliminary analytic evaluation of the 
evaporation rate 
within the framework of Horizon Quantum Mechanics and some  considerations related to the conservation of energy, 
suggest that the most plausible picture is the
one with a negative deformation parameter, $\beta<0$.

In this connection, we emphasize that
a similar result would not be surprising 
in the context of a corpuscular (i.e discrete) description
of black holes; in Ref.~\cite{Jizba:2009qf}, indeed,
it was shown that a GUP with $\beta<0$
can be derived assuming that the universe
has an underlying crystal lattice-like structure. 
Moreover, in Ref.~\cite{Ong:2018zqn} it was
found that $\beta<0$ is the only choice
compatible with the Chandrasekhar limit; in other terms, 
the GUP with positive deformation parameter 
would allow arbitrarily large white dwarfs to exist, 
a result that clashes with current astrophysical observations.
In Section~\ref{uncert-sec} we have also seen that, in the case
$\beta>0$, black holes would not
entirely evaporate. However, if on the one hand
black hole remnants may be viewed as 
potential candidates for dark matter~\cite{Chen:2004rha},
on the other hand their existence would be rather problematic
for the entropy/information
paradox~\cite{Bekenstein:1993dz}.

Apart from these supporting arguments, it also worth 
remarking that,
if $\beta<0$, then from Eq.~\eqref{gup} 
there would be a maximum value of
$\delta p$ around the Planck scale 
for which $\delta x\hspace{0.3mm}\delta p=0$, 
i.e. the quantum uncertainty would be completely
erased. Therefore, it would seem that
physics would can become classical again
at Planck scale. The possibility of a 
classical Planckian regime has been
already addressed in Ref.~\cite{Hossenfelder:2012uy}, 
by regarding $\hbar$ as 
a dynamical field that
vanishes in the Planckian limit.

Given the absolute lack of knowledge
about physics at Planck scale, it is clear that in principle 
all possible scenarios should be contemplated
in order to achieve a better understanding of
how gravity and quantum effects behave
when combined together. For example, one cannot
exclude a priori the possibility that $\beta$ is a
function rather than a pure number.
This has been originally suggested 
in Ref.~\cite{Chen:2014bva}
as condition for black hole complementarity principle
to always hold. 
A similar result has been recently recovered in Ref.~\cite{Vagenas:2018zoz}, 
where, by conjecturing the equality between 
the GUP--deformed black hole temperature
of a Schwarzschild black hole and the modified Hawking temperature of a quantum-corrected
Schwarzschild black hole, it has been obtained
a GUP parameter depending on the ratio $m_p/M$.

In light of the above discussion, the present
investigation should thus be regarded as a further attempt to
gain information about the 
GUP black hole physics 
through a connection with an intrinsically 
corpuscular description of gravity.
More work is inevitably required 
to provide a definite answer
about the sign of the GUP deformation parameter 
$\beta$~\cite{inprepa}.


\acknowledgments
The authors would like to thank Roberto Casadio and Andrea Giusti for helpful conversations.


\begin{thebibliography}{0}
	
	\bibitem{aHeisenberg:1927zz} 
	W.~ Heisenberg,
	Z.\ Phys.\  {\bf 43}, 172 (1927).
	
	\bibitem{Maggiore:1993zu} 
  M.~Maggiore,
  Phys.\ Rev.\ D {\bf 49}, 5182 (1994).
  
  \bibitem{Wheeler:1955zz} 
  J.~A.~Wheeler,
  Phys.\ Rev.\  {\bf 97}, 511 (1955).
	
	\bibitem{Snyder:1946qz} 
	H.~S.~Snyder,
	Phys.\ Rev.\  {\bf 71}, 38 (1947).
	
	\bibitem{Yang:1947ud} 
	C.~N.~Yang,
	Phys.\ Rev.\  {\bf 72}, 874 (1947).
	
	\bibitem{maed}C.A. Mead, Phys. Rev. 135, B849 (1964).
	
	\bibitem{Karolyhazy:1966zz} 
	F.~Karolyhazy,
	Nuovo Cim.\ A {\bf 42}, 390 (1966).
	
	\bibitem{Amati:1987wq} 
	D.~Amati, M.~Ciafaloni and G.~Veneziano,
	Phys.\ Lett.\ B {\bf 197}, 81 (1987).
	
	\bibitem{Gross:1987kza} 
	D.~J.~Gross and P.~F.~Mende,
	Phys.\ Lett.\ B {\bf 197}, 129 (1987).
	
	\bibitem{Amati:1988tn} 
	D.~Amati, M.~Ciafaloni and G.~Veneziano,
	Phys.\ Lett.\ B {\bf 216}, 41 (1989).
	
	\bibitem{Konishi:1989wk} 
	K.~Konishi, G.~Paffuti and P.~Provero,
	Phys.\ Lett.\ B {\bf 234}, 276 (1990).
	
	\bibitem{Maggiore:1993kv} 
	M.~Maggiore,
	Phys.\ Lett.\ B {\bf 319}, 83 (1993).
	
	\bibitem{Scardigli:1995qd} 
	F.~Scardigli,
	Nuovo Cim.\ B {\bf 110}, 1029 (1995). 
	
	\bibitem{FS}
	F.~Scardigli,
	Phys.\ Lett.\ B {\bf 452}, 39 (1999).
	
	\bibitem{Capozziello:1999wx} 
	S.~Capozziello, G.~Lambiase and G.~Scarpetta,
	Int.\ J.\ Theor.\ Phys.\  {\bf 39}, 15 (2000).
	
	\bibitem{Bojowald:2011jd} 
	M.~Bojowald and A.~Kempf,
	Phys.\ Rev.\ D {\bf 86}, 085017 (2012).
	
	\bibitem{Scardigli:2003kr} 
	F.~Scardigli and R.~Casadio,
	Class.\ Quant.\ Grav.\  {\bf 20}, 3915 (2003).
	
	
	
	\bibitem{Adler:1999bu} 
	R.~J.~Adler and D.~I.~Santiago,
	Mod.\ Phys.\ Lett.\ A {\bf 14}, 1371 (1999).
	
	  \bibitem{Adler}
	R.~J.~Adler, P.~Chen and D.~I.~Santiago,
	Gen.\ Rel.\ Grav.\  {\bf 33}, 2101 (2001).
	
	\bibitem{Casadio:2009jc} 
	R.~Casadio, R.~Garattini and F.~Scardigli,
	Phys.\ Lett.\ B {\bf 679}, 156 (2009).
	
	\bibitem{Casadio:2013aua} 
	R.~Casadio and F.~Scardigli,
	Eur.\ Phys.\ J.\ C {\bf 74}, 2685 (2014).
	
	\bibitem{Scardigli:2016pjs} 
	F.~Scardigli, G.~Lambiase and E.~Vagenas,
	Phys.\ Lett.\ B {\bf 767}, 242 (2017).
	
	\bibitem{VAF}
	E.~C.~Vagenas, S.~M.~Alsaleh and A.~Farag,
	EPL {\bf 120},  40001 (2017).
	
	\bibitem{Lambiase:2017adh} 
	G.~Lambiase and F.~Scardigli,
	Phys.\ Rev.\ D {\bf 97},  075003 (2018).
	
	  \bibitem{Scardigli:2018jlm} 
	F.~Scardigli, M.~Blasone, G.~Luciano and R.~Casadio,
	Eur.\ Phys.\ J.\ C {\bf 78},  728 (2018).
	
\bibitem{Luciano2019} 
  G.~G.~Luciano and L.~Petruzziello,
  arXiv:1902.07059 [hep-th].
  
    \bibitem{Kanazawa:2019llj} 
  T.~Kanazawa, G.~Lambiase, G.~Vilasi and A.~Yoshioka,
  Eur.\ Phys.\ J.\ C {\bf 79},  95 (2019).
  
    \bibitem{Jizba:2009qf} 
  P.~Jizba, H.~Kleinert and F.~Scardigli,
  Phys.\ Rev.\ D {\bf 81}, 084030 (2010).
  
   \bibitem{Ong:2018zqn}
 Y.~C.~Ong,
 JCAP {\bf 1809}, 015  (2018).
 
   \bibitem{Dvali:2011aa} 
  G.~Dvali and C.~Gomez,
  Fortsch.\ Phys.\  {\bf 61}, 742 (2013). 

  
  \bibitem{Dvali:2012en} 
  G.~Dvali and C.~Gomez,
  Eur.\ Phys.\ J.\ C {\bf 74}, 2752 (2014).
  
  \bibitem{Casadio:2013ulk} 
  R.~Casadio and A.~Orlandi,
  JHEP {\bf 1308}, 025 (2013).
  
    \bibitem{Casadio:2014vja} 
  R.~Casadio, A.~Giugno, O.~Micu and A.~Orlandi,
  Phys.\ Rev.\ D {\bf 90},  084040 (2014).
  
  \bibitem{Dvali:2013vxa} 
  G.~Dvali, D.~Flassig, C.~Gomez, A.~Pritzel and N.~Wintergerst,
  Phys.\ Rev.\ D {\bf 88},  124041 (2013).
  
  \bibitem{Dvali:2013eja} 
  G.~Dvali and C.~Gomez,
  JCAP {\bf 1401}, 023 (2014).
  
  \bibitem{Dvali:2014ila} 
  G.~Dvali, C.~Gomez, R.~S.~Isermann, D.~Lüst and S.~Stieberger,
  Nucl.\ Phys.\ B {\bf 893}, 187 (2015).
  
  \bibitem{Dvali:2015aja} 
  G.~Dvali,
  Fortsch.\ Phys.\  {\bf 64}, 106 (2016).
  
  
  \bibitem{Casadio:2015bna} 
  R.~Casadio, A.~Giugno and A.~Orlandi,
  Phys.\ Rev.\ D {\bf 91}, 124069 (2015).
  
  \bibitem{Casadio:2016zpl} 
  R.~Casadio, A.~Giugno and A.~Giusti,
  Phys.\ Lett.\ B {\bf 763}, 337 (2016).
  
  \bibitem{Casadio:2016aum} 
  R.~Casadio and R.~da Rocha,
  Phys.\ Lett.\ B {\bf 763}, 434 (2016).
  
  \bibitem{giusti-tesi}
  A. Giusti, ``On the Corpuscular Theory of Gravity'' Int. J. Geom. Meth. Mod. Phys. (to
  appear) (2019),  doi:10.1142/S0219887819300010.
  
  
  
  \bibitem{Page:1976df} 
  D.~N.~Page,
  Phys.\ Rev.\ D {\bf 13}, 198 (1976).
  
  
  \bibitem{Aharonov:1987tp} 
  Y.~Aharonov, A.~Casher and S.~Nussinov,
  Phys.\ Lett.\ B {\bf 191}, 51 (1987).
  
  
  \bibitem{Mathur:2009hf} 
  S.~D.~Mathur,
  Class.\ Quant.\ Grav.\  {\bf 26}, 224001 (2009).
  
  \bibitem{Bekenstein:1993dz} 
  J.~D.~Bekenstein,
  Phys.\ Rev.\ D {\bf 49}, 1912 (1994).
  
  \bibitem{hawk}
  S.~W.~Hawking and G.~F.~R.~Ellis,
  \emph{The Large Scale Structure of Space-Time}, 
  Cambridge University Press, Cambridge (2011).
  
      \bibitem{Dvali:2018tno} 
  G.~Dvali,
  Found.\ Phys.\  {\bf 48}, 1219 (2018).
  
  \bibitem{Bogolyubov:1947zz} 
  N.~N.~Bogolyubov,
  J.\ Phys.\ (USSR) {\bf 11}, 23 (1947)
  [Izv.\ Akad.\ Nauk Ser.\ Fiz.\  {\bf 11}, 77 (1947)].
  

  
  \bibitem{horizon-QM}
   R.~Casadio,
  arXiv:1305.3195 [gr-qc];
  R.~Casadio, A.~Giugno and A.~Giusti,
  Gen.\ Rel.\ Grav.\  {\bf 49},  32 (2017)
  
  \bibitem{Casadio:1997yv} 
  R.~Casadio, B.~Harms and Y.~Leblanc,
  Phys.\ Rev.\ D {\bf 58}, 044014 (1998).
  
  \bibitem{Alberghi:2006km} 
  G.~L.~Alberghi, R.~Casadio and A.~Tronconi,
  J.\ Phys.\ G {\bf 34}, 767 (2007).
  
    \bibitem{Chen:2004rha} 
  P.~Chen,
  Mod.\ Phys.\ Lett.\ A {\bf 19}, 1047 (2004);
  New Astron.\ Rev.\  {\bf 49}, 233 (2005).
  
  \bibitem{Hossenfelder:2012uy} 
  S.~Hossenfelder,
  Phys.\ Lett.\ B {\bf 725}, 473 (2013).
  
  
\bibitem{Chen:2014bva} 
  P.~Chen, Y.~C.~Ong and D.~h.~Yeom,
  JHEP {\bf 1412}, 021 (2014).  

\bibitem{Vagenas:2018zoz} 
  E.~C.~Vagenas, S.~M.~Alsaleh and A.~Farag,
  EPL {\bf 120}, 40001 (2017).
  
  \bibitem{inprepa}
  L.~Buoninfante, G.~G.~Luciano and L.~Petruzziello, 
  \emph{in preparation}.

\end{thebibliography}
\end{document}